\documentclass{camera}
\usepackage{graphicx}  

\begin{document}

\title{Fluctuations of the largest fragment size in percolation and
multifragmentation}

\author{J. Brzychczyk$^{1}$, T. Pietrzak$^{1}$, A. Wieloch$^{1}$
\and W. Trautmann$^{2}$}

%
\organization{
$^{1}$ Institute of Physics, Jagiellonian University, 30-059 Krak\'{o}w, Poland\\
$^{2}$ GSI Darmstadt, D-64291 Darmstadt, Germany}

\maketitle

\begin{abstract}
Aladin data on fragmentation of $^{197}$Au projectiles
are remarkably well reproduced by a bond percolation model.
A critical behavior is identified on the basis of fluctuations
of the largest fragment size.
\end{abstract}

The present work is motivated by percolation studies \cite{jb}
which demonstrated that a cumulant analysis of
the largest fragment size distributions is a valuable tool
in searching for a phase transition (critical behavior)
in fragmenting systems.
This method is applicable even to very small systems
and therefore is well suited for applications
to nuclear multifragmentation.

The simulations were performed
with a three-dimensional bond percolation model on simple cubic
lattices with free boundary conditions to account for
the surface presence.
Given a bond probability value $p$ (control parameter) and 
the total number of sites $Z_{0}$ (system size),
the probability distribution $P(Z_{max})$
of the largest cluster size $Z_{max}$ is determined from
a large sample of events.
The statistical measures as the mean, variance,
skewness and kurtosis contain the most significant
information about the distribution.
Of particular interest are the following cumulant ratios
\begin{eqnarray}
K_2\equiv &\mu_{2}/\langle Z_{max}\rangle^2& =\kappa_2/\kappa_1^2 \nonumber\\
K_3\equiv &\mu_{3}/\mu_{2}^{3/2}& =\kappa_3/\kappa_2^{3/2} \nonumber\\
K_4\equiv &\mu_{4}/\mu_{2}^{2}-3& =\kappa_4/\kappa_2^{2},
\end{eqnarray}
where $\langle Z_{max}\rangle$ denotes the mean value,
$\mu_i=\langle(Z_{max}-\langle Z_{max}\rangle)^i\rangle$ is the $i$th
central moment, and $\kappa_i$ is the $i$th cumulant of $P(Z_{max})$.
$K_{2}$ is the variance normalized to the squared mean,
$K_{3}$ is the skewness which indicates the distribution asymmetry, and
$K_{4}$ is the kurtosis excess measuring the degree of peakedness
with respect to the normal distribution.
In the transition region, these quantities obey with good accuracy
finite-size scaling relations even for very small systems
with open boundaries. This allows to identify universal
(independent of the system size) features of $K_{i}$ at the
percolation transition.
The transition in finite systems (the pseudocritical point)
is associated with the broadest and most symmetric $P(Z_{max})$ distribution,
which is indicated by
$K_{3}=0$ and the minimum value of $K_{4}$ of about $-1$.
This criticality signal is approximately
preserved when events are sorted
by measurable variables correlated with
the control parameter ({\it e.g.} $Z_{bound}$) \cite{jb}.

\begin{figure}[ht]
\begin{center}
\includegraphics[width=10.0cm]{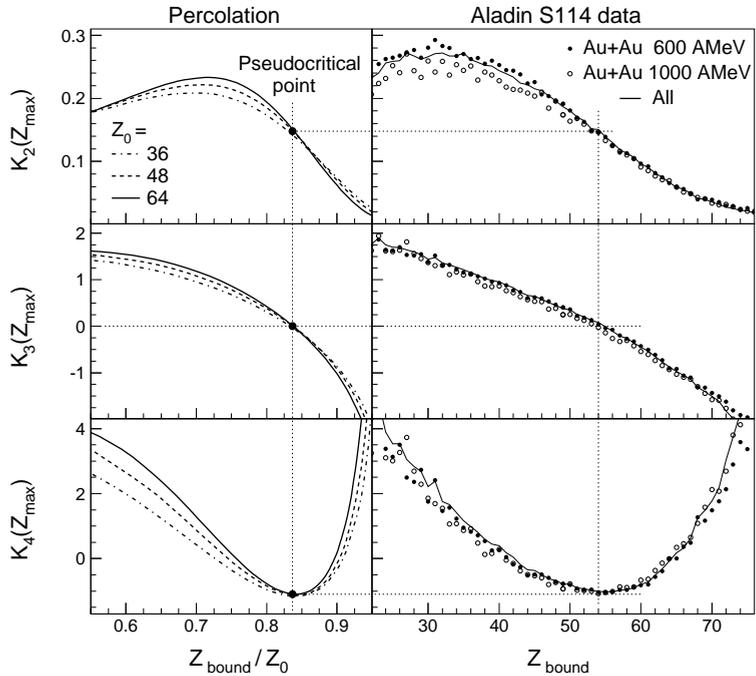}
\caption{The cumulant ratios $K_{i}$ of the $P(Z_{max})$ distribution as a function of $Z_{bound}$.
Bond percolation calculations versus the experimental data.}
\label{fig01} 
\end{center}
\end{figure}

The present work compares percolation predictions with
the Aladin S114 data on fragmentation of projectile
spectators in $^{197}$Au + Au, In, Cu collisions at
the incident energies of 600-1000 AMeV.
Details of the experiment and general
characteristics of the data were presented in \cite{sch}.

Figure 1 examines the cumulants ratios $K_{i}$ of the largest fragment
size distribution $P(Z_{max})$. The percolation results are plotted
in the left diagrams as a function of $Z_{bound}$ normalized to the
system size $Z_{0}$ for three different system sizes that span
over a range expected in the transition region.
Percolation events were generated for the bond probabilities
uniformly distributed in the interval $[0,1]$, and then sorted
according to $Z_{bound}$. As one can see,
the pseudocritical point is located at 
$Z_{bound}/Z_{0} = 0.84$ independently of the system size.
The experimental results are shown in the right diagrams for
the Au + Au systems at 600 and 1000 AMeV
and for the summed data sets (all targets and energies).
Here, $K_{i}$ are plotted as a function of $Z_{bound}$.
The percolation and experimental patterns of the cumulants
are very similar. In particular, the percolation pseudocritical
point is well reflected in the data at $Z_{bound} = 54$.
Based on this comparison, the (mean) system size at $Z_{bound} = 54$
can be estimated as $Z_{0} = Z_{bound}/0.84\simeq 64$.

\begin{figure}[ht]
\begin{center}
\includegraphics[width=8.0cm]{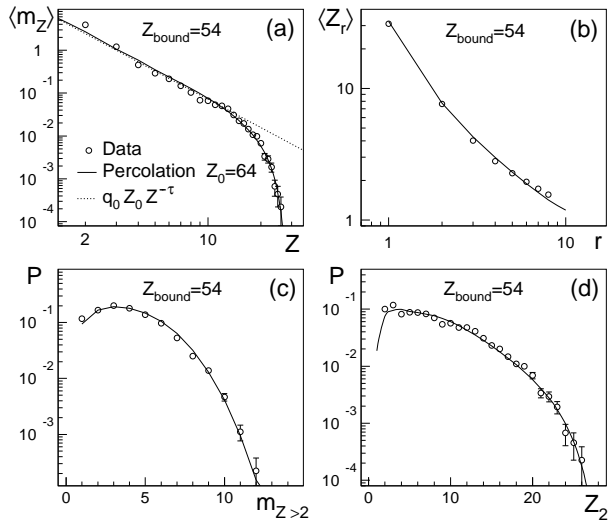}
\caption{Percolation predictions versus the experimental data
at $Z_{bound} = 54$: (a) the mean fragment multiplicity as a
function of the fragment size (the largest fragment excluded),
(b) the mean fragment size as a function of the fragment rank,
(c) the multiplicity distribution of fragments with $Z>2$,
(d) the distribution of the second largest fragment charge.}
\label{fig02} 
\end{center}
\end{figure}

Once the system size is established,
we can examine other observables related
to fragment charge partitions. In Fig. 2 we compare percolation
results with the data at the pseudocritical point $Z_{bound} = 54$.
Panel (a) shows the fragment size distribution.
The model well describes the data over four orders
of magnitude. As expected for the percolation
pseudocritical point, the distribution follows in some range
the asymptotic power-law dependence shown by the dotted line \cite{jb}.
Panel (b) shows the Zipf-type plot, {\it i.e.} the mean size
of the largest, second largest,... $r$-largest fragments plotted
against their rank $r$.
The percolation model very well reproduces not only mean values
but also event-to-event fluctuations.
For example, the next panels show the multiplicity distribution
of fragments with $Z>2$ and the distribution of the second largest fragment charge.

Similar comparisons performed at other $Z_{bound}$
values in the range from 36 to 66 have also shown
an almost perfect resemblance between percolation predictions and the data.
The system sizes determined in the analysis
are in a good agreement with the experimental estimates \cite{poch}.

In conclusions,
fluctuations of the largest fragment charge observed in the $^{197}$Au
spectator fragmentations show the percolation pattern.
In analogy to percolation, the pseudocritical point is
identified at $Z_{bound} = 54$ which corresponds to the He-Li isotope temperature
corrected for secondary decays of $5.2 \pm 0.4$ MeV \cite{traut}.
Detailed comparisons have demonstrated that
the experimental fragment charge partitions are remarkably well reproduced by the bond percolation
model with no free parameters and no corrections for
secondary decays in a wide range of $Z _{bound}$.
In the context of the lattice gas model which is equivalent to bond percolation
at the normal density, the success of percolation suggests that
clusters are formed at the dense medium and isolated fragments are cold,
in line with the ''Little big bang" scenario of multifragmentation
\cite{campi}.

\smallskip

This work has been supported by the Polish Ministry
of Science and Higher Education grant N202 160 32/4308 (2007-2009).

\end{document}